\documentclass[sigconf]{acmart}
\AtBeginDocument{%
  }

\copyrightyear{2026}
\acmYear{2026}
\setcopyright{cc}
\setcctype{by-nc-nd}
\acmConference[CHI '26]{Proceedings of the 2026 CHI Conference on Human Factors in Computing Systems}{April 13--17, 2026}{Barcelona, Spain}
\acmBooktitle{Proceedings of the 2026 CHI Conference on Human Factors in Computing Systems (CHI '26), April 13--17, 2026, Barcelona, Spain}
\acmPrice{}
\acmDOI{10.1145/3772318.3790639}
\acmISBN{979-8-4007-2278-3/2026/04}

\usepackage{array}
\newcolumntype{L}[1]{>{\raggedright\let\newline\\\arraybackslash\hspace{0pt}}m{#1}}
\newcolumntype{C}[1]{>{\centering\let\newline\\\arraybackslash\hspace{0pt}}m{#1}}
\newcolumntype{R}[1]{>{\raggedleft\let\newline\\\arraybackslash\hspace{0pt}}m{#1}}
\usepackage{colortbl} 
\definecolor{mypink2}{RGB}{239, 240, 255}

\begin{document}

\title{\emph{The Words That Can't Be Shared}: Exploring the Design of Unsent Messages}


\author{Michael Yin}
\orcid{0000-0003-1164-5229}
\affiliation{
  \institution{University of British Columbia}
  \city{Vancouver}
  \state{BC}
  \country{Canada}
  \postcode{V6T 1Z4}
}
\email{jiyin@cs.ubc.ca}

\author{Robert Xiao}
\orcid{0000-0003-4306-8825}
\affiliation{
  \institution{University of British Columbia}
  \city{Vancouver}
  \state{BC}
  \country{Canada}
  \postcode{V6T 1Z4}
}
\email{brx@cs.ubc.ca}


\begin{abstract}
People often have things they want to say but hold back in conversations, fearing being vulnerable or facing social consequences. Online, this restraint can take a distinctive form: even when such thoughts are written out --- in moments of anger, guilt, or longing --- people may choose to withhold them, leaving them \emph{unsent}. This process is underexamined; we investigate the experience of writing such messages within people's digital communications. We find that unsent messages become expressive containers for suppressed feelings, where the act of writing creates a pause for reflection on the relationship and oneself. Building on these insights, we probed into how the design of the writing platforms of unsent messages affects people's experiences and motivations. Speculating with participants on nine evocative variants of a note-taking platform, we highlight how design shapes the emotional, temporal, and ritualistic qualities of unsent messages, revealing tensions between people's social desires and communicative actions. 
\end{abstract}


\maketitle

\section{Introduction}

Suppressing what we want to say to others is a universal experience for all humans. People may hold strong feelings towards others, such as anger, guilt, or regret, yet hesitate in expressing these feelings to the other person in fear of seeming overly vulnerable, harming their relationship, or giving rise to repercussions that they do not wish to bear \cite{andalibiDisclosurePrivacyStigma2020, andalibiTestingWatersSending2018, bar-talSelfCensorshipSocioPoliticalPsychologicalPhenomenon2017}. One way of containing these strong, directed feelings is to translate them into unsent letters --- written messages later discarded or left unseen. This process is not new: Abraham Lincoln reportedly wrote such letters to his contemporaries, putting his feelings into notes until his emotions subsided\footnote{\url{https://www.nytimes.com/2014/03/23/opinion/sunday/the-lost-art-of-the-unsent-angry-letter.html}}. Although the shape and form of writing such letters has changed in the modern day, the practice persists --- evidenced by how people write such messages on platforms such as the Unsent Letters subreddit\footnote{\url{https://www.reddit.com/r/UnsentLetters/}} or The Unsent Project website\footnote{\url{https://theunsentproject.com/}}. In doing so, people can \emph{externalize} their feelings towards others without necessarily needing to \emph{share} them. 
 
Still, sharing and disclosure are important relational processes. In the increasingly digital landscape, people share for self-approval, intimacy, or distress relief \cite{vitakYouCantBlock2014}; disclosure is important for building belonging, connection, and friendship \cite{rubinFriendshipProximitySelfdisclosure1978, davisFriendship20Adolescents2012, veltmanAristotleKantSelfDisclosure2004}. Yet, while disclosing feelings can build relationships, withholding thoughts can preserve them. This phenomenon, sometimes coined \emph{non-disclosure} or \emph{self-censorship}, can arise due to the interpersonal, self-perception, and impression-based risks of sharing \cite{vitakYouCantBlock2014}. Even as people hold strong personal thoughts and desires, expressing these to others may cause harm and disrupt harmony, while not even leading to the intended goal \cite{andalibiDisclosurePrivacyStigma2020, robertsNobodySaysNo2011, bar-talSelfCensorshipSocioPoliticalPsychologicalPhenomenon2017}. To handle this tension between hidden wants and outward actions, people may vent out their frustrations in their private worlds \cite{robertsNobodySaysNo2011}. In this work, we explore unsent messages as a representation of this private space --- a space between desire and decision, giving form to innate feelings without public expression. 

Despite the near-universal experience of and motivations underlying unsent messages, there has been little academic research exploring it as a digital practice. Situated in a liminal space between feelings and action, unsent messages are neither entirely private like a journal or diary, nor fully social like a sent message. We explore how, why, and in what contexts people feel compelled to write unsent messages. We find that people crave connection, but are also concerned about the receiver's perception and their own self-presentation. We use these findings to explore the design of platforms on which people write unsent messages, and assess how speculative designs might affect the overarching practice. 

Overall, we explore the two research questions of:

\begin{itemize}
    \item \textbf{RQ1}: What are the motivations, written contents, and emotional outcomes of unsent messages?
    \item \textbf{RQ2}: How does the design of digital writing platforms affect the experience of unsent messages?
\end{itemize}

To address the first research question, we conducted a formative study with participants who had experience writing such unsent messages. Through listening to their stories, we drew out insights about their motivations, feelings, and reflections, understanding how unsent messages represent \textbf{reflective containers for people's feelings that they feel uncomfortable with expressing}. To handle the tension between strongly felt emotions towards another and the perceived risk of communicating these emotions, people use unsent messages as a flexible, personal space for various invisible outcomes and goals --- rehearsing, understanding, expressing, holding onto, or letting go of their feelings.

Most commonly, participants defaulted to a ``Notes'' app\footnote{often the default note-taking app on a smartphone} as an improvised manifestation of this space. Our subsequent study examines how \emph{intentional design} of this space can support and modulate different reflective outcomes. To do so, we built nine variants of a Notes app; each variant represents a lightweight design probe that experiments with dimensions of rituality, temporality, sociality, and affordances. Through speculating with participants using these probes, we address the second research question: exploring \textbf{how design can influence how people contain, express, and let go of their feelings} and \textbf{make certain emotional goals feel more difficult or more attainable}. 

Our work contributes (1) an empirical understanding of the motivations and experiences of unsent messages, a phenomenon presently understudied in HCI, (2) an exploration of the design space of unsent messages through a design probe, and (3) implications and suggestions for how designers and researchers can intentionally design reflective spaces to support specific emotional outcomes. 

\section{Background and Related Work}

We first consider the nature of online disclosure, understanding why people feel motivated to share or withhold their thoughts and feelings. We then contextualize disclosure around computer-mediated communication (CMC) to understand how theory underpins motivations of disclosure and withholding. We consider the relationship between writing and management of difficult human emotions, and finally spotlight ways in which HCI work has encouraged reflection and growth, grounded in human experiences and feelings. 

\subsection{Online Expression and Self-Censorship}

Sharing and disclosure are important --- people have a deep-seated drive to share \cite{bar-talSelfCensorshipSocioPoliticalPsychologicalPhenomenon2017}, which bridges belonging and closeness in a relationship \cite{davisFriendship20Adolescents2012}. Suppression of thoughts and emotions requires cognitive effort, which can create psychological preoccupation \cite{wegnerParadoxicalEffectsThought1987}; on the other hand, online self-disclosure can promote positive well-being through potentially liberating factors of connectedness, social support, and psychological authenticity \cite{luoSelfdisclosureSocialMedia2020, andalibiWhatHappensDisclosing2019}. 

Conversely, what factors give rise to the opposite --- of \emph{non-disclosure} or \emph{self-censorship}? Many researchers have investigated how factors regarding self-image and an imagined audience can lead to such outcomes \cite{andalibiDisclosurePrivacyStigma2020}; divulgence of certain information can violate existing principles or norms, which can in turn lead to harm \cite{bar-talSelfCensorshipSocioPoliticalPsychologicalPhenomenon2017}. Social media users, for instance, may self-censor because they want to avoid arguments and adverse reactions, while maintaining their personal reputation and self-presentation \cite{madsenSelfcensorshipInternalSocial2016, sleeperPostThatWasnt2013, zhangPoliticallyMotivatedRetroactive, vitakYouCantBlock2014}. Mismanaging online self-presentation can have real-life consequences, such as social backlash and rejection \cite{huWorkProgressGlance2023, gagrcinYourSocialTies2024}. While not all thoughts might be appropriate to share in all scenarios \cite{kennedyOversharingNorm2018, shabahangOversharingSocialMedia2024}, depending on a subjective reading of social context and boundaries, self-censorship might come at a cost of repressing one's true feelings. For instance, Roberts and Nason \cite{robertsNobodySaysNo2011} found that students' negative feelings towards others were avoided in a public forum but laid bare in a private one. 

Andalibi \cite{andalibiDisclosurePrivacyStigma2020} investigated the dimensions that mediate digital disclosure and non-disclosure. The familial norms people grow up with, their openness to failure, and their anticipation of reactions to disclosure all have determining roles in this process \cite{andalibiDisclosurePrivacyStigma2020}. Das and Kramer highlighted how non-disclosure is more likely when people find it difficult to define their audience \cite{dasSelfCensorshipFacebook2013}. In general, it can be hard for people to understand what is appropriate to share online and to whom \cite{liuBuildingPersonalizedModel2022}. Furthermore, even as people chase feelings of support, acknowledgement, and catharsis \cite{andalibiWhatHappensDisclosing2019, andalibiDisclosurePrivacyStigma2020, andalibiSocialSupportReciprocity2018}, they may fear vulnerability and loss of control over their information \cite{andalibiDisclosurePrivacyStigma2020}. Jordan highlights how it takes courage for people to address vulnerability and potentially move into conflict \cite{jordan1990courage, jordanValuingVulnerabilityNew2008}; in balancing this tension, people do not always decide to speak up. In a relationship, one party might employ avoidance of the other \cite{roloffConflictManagementAvoidance1999, scissorsRoomInterpretationRole2014}, which can be more easily achieved through digital communication \cite{scissorsRoomInterpretationRole2014}. Also, people might adapt indirect disclosure processes \cite{andalibiTestingWatersSending2018} such as subtle hints or expressive artwork, to protect themselves from risk while still satisfying their psychological and social needs.

However, disclosure is not solely tied to human psychology, but also the affordances and design of the communication platform \cite{ichinoIveTalkedIntending2022, andalibiDisclosurePrivacyStigma2020}. Ichino et al. \cite{ichinoIveTalkedIntending2022} found that disclosure was encouraged in a system that represented humans as avatars distinct from their human appearance, suggesting an inverse relationship between disclosure and literal self-representation. Furthermore, social media platforms such as Facebook may be perceived as too public to engage in discussion about sensitive topics without risk \cite{gurgunWhyWeNot2024}. We consider the specific representation of non-disclosure of unsent messages --- their construction, their purpose, and their impact on the relational context. Although much prior work has viewed disclosure and self-censorship through public social media, similar concepts can contextualize motivations in more intimate conversations --- the disinclination to harm, the weighing of risk, and the extended social dynamics. 

\subsection{Computer-Mediated Communication (CMC) in One-to-One Messaging}

The contemporary form of unsent messages is strongly tied to the affordances and constraints of computer-mediated communication (CMC). Early CMC theories tended to fall under the \emph{cues-filtered-out} grouping, highlighting the lack of nonverbal cues when compared to face-to-face (FtF) \cite{walther2011theories}. While some pointed to consequences such as lack of social presence or context clues, Walther's hyperpersonal model \cite{waltherComputerMediatedCommunicationImpersonal1996d, waltherLanguagePsychologyNew2021} hypothesized that CMC can surpass FtF in terms of emotion and affection. Two distinct relational factors of this model are the idealized perception of the receiver, in which people might overattribute certain qualities to the other party, and an optimization of self-presentation, in which people perfect their response for a favourable impression. The initial hyperpersonal model has been tested and commented on through the years \cite{rainsExaminingQualitySocial2019, scottDoesRecentResearch2020, waltherSelectiveSelfpresentationComputermediated2007}, adding increased context and nuance. For instance, Rains et al. showed that \emph{removing} social cues would \emph{increase} motivations for participants to aid a confederate \cite{rainsExaminingQualitySocial2019}. Furthermore, even without expression cues, communicators adapt to other affordances \cite{rabbyComputerMediatedCommunicationEffects2003} such as paralinguistic cues in punctuation and emojis \cite{liebman2016, sidiYouGetWhat2021}. 

The asynchronicity of CMC further affects how people communicate. With the different temporal scale of CMC, people hold greater control over their interactions --- they can rehearse their constructed messages before sending them out, allowing them to optimize their effects \cite{waltherComputerMediatedCommunicationImpersonal1996d, waltherSelectiveSelfpresentationComputermediated2007}. This has implications for how people construct, edit, and even withhold messages. With non-disclosure, this temporal difference also allows for disappearance, which happens when people no longer respond. This may arise when one feels difficulty in communicating their feelings without confrontation \cite{thomasDisappearingAgeHypervisibility2021b}, even when such difficult conversations also develop intimacy and trust.

Personal desires, self-presentation, and anticipated audience reaction all play a role in digital communication dynamics, both in terms of messages sent and feelings unsaid \cite{managoSelfpresentationGenderMySpace2008, andalibiDisclosurePrivacyStigma2020, andalibiTestingWatersSending2018, sleeperPostThatWasnt2013}. Tying into a dramaturgical perspective on CMC \cite{hoganPresentationSelfAge2010}, users constantly refine how others perceive them. How do users maintain and construct self-image, and how does that affect their message construction, perfection, and non-divulgence? We consider how self-presentation and social imaginaries mediate both people's writings of messages and their decisions to leave them unsent. 

\subsection{Writing, Self-Reflection, and Emotional Processing}

Documenting insights and experiences in writing allows people to learn about themselves and the broader world \cite{walkerWritingReflection1985}. One way in which such reflective writing manifests is through journaling, the most extensively reviewed practice in the literature. Journaling has many benefits in terms of personal reflection --- Cowan argues that journaling, as a conversation with self, helps people look both backwards towards past learning and forward towards future challenges \cite{cowanFacilitatingReflectiveJournalling2013a}. Journaling helps people understand and conceptualize their experiences \cite{sendallJournallingPublicHealth2013a,hubbsPaperMirrorUnderstanding2005, bleicherUsingSmallMoments2011, newtonLearningReflectJourney2004}, potentially spurring growth through psychological and emotional benefits \cite{ullrichJournalingStressfulEvents2002c, eppValueReflectiveJournaling2008, utleyTherapeuticUseJournaling2011a}. Fritson considers how journaling affects one's self-efficacy and locus of control, affecting personal agency over one's actions \cite{fritsonImpactJournalingStudents2008a}. As a result, journaling supports moving along the levels of reflection \cite{fleckReflectingReflectionFraming2010}, as users ponder about their experiences, explore relationships, and consider future changes. While journaling in general already supports positive effects, Pennebaker espoused \emph{expressive writing} \cite{pennebakerWritingEmotionalExperiences1997a, pennebakerExpressiveWritingPsychological2018}, a more structured form of writing in which authors are invited to explore their most emotional moments; such writing can create short-term distress but invoke long-term psychological and health benefits \cite{baikieEmotionalPhysicalHealth2005a, pennebakerWritingEmotionalExperiences1997a, gortnerBenefitsExpressiveWriting2006a}. 

In HCI research, Lin et al. considered how the ecosystem and design of written paper journaling affect its practice \cite{linCraftingPersonalJournaling2025}. Journaling is affected by how it is conducted, and much prior work has focused on extending the practice into a digital form, i.e. through smart journals \cite{elsdenItsJustMy2016}. Jang et al.'s EmotionFrame \cite{jangDesignFieldTrial2024} and Karaturhan et al.'s digital journaling app \cite{karaturhanCombiningMomentaryRetrospective2022b} both support digitally archiving personal experiences beyond the boundaries of writing. Similarly, Isaacs et al.'s \emph{Echo} \cite{isaacsEchoesHowTechnology2013a} highlights how archiving and reflecting on life events can promote well-being and future learning. 

More recently, generative AI has also been explored in supporting journaling \cite{nepalContextualAIJournaling2024, nepalMindScapeStudyIntegrating2024b, zhouJournalAIdeEmpoweringOlder2025a, kimDiaryMateUnderstandingUser2024b, angeniusInteractiveJournalingAI2023}. Artificial conversational agents offer a non-judgmental space for expression and reflection \cite{angeniusInteractiveJournalingAI2023}, for instance, JournalAIde \cite{zhouJournalAIdeEmpoweringOlder2025a} and DiaryMate \cite{kimDiaryMateUnderstandingUser2024b} both offer LLM-based conversational agents. Across AI-guided journaling and reflection literature, conversational agents can provide reflective and social assistance, offering ways of managing mental health to prevent people from entering a ruminative cycle \cite{jungMyListenerAIMediatedJournaling2024a, kimReflectiveAgencyEthical2025, kimDiaryMateUnderstandingUser2024b}. Even as such systems offer empathy, comfort, and emotional regulation, they also introduce key considerations regarding tradeoffs in control and agency during reflective practice \cite{kimDiaryMateUnderstandingUser2024b, jungMyListenerAIMediatedJournaling2024a}. 

Unsent messages are not the same as journaling --- journaling has often been studied in a more structured setting, whereas unsent messages may be more impromptu and interpersonal. Furthermore, Roper argues that unsent letters straddle the boundary between self and others, lacking the ``artifice'' of personal writing \cite{roperSplittingUnsentLetters2001}. Still, considering the lack of research on unsent messages and how unsent messages and journaling are both writing practices that document experience and thoughts, we draw comparisons and distinctions in terms of these reflective practices. 

\subsection{Designing Reflective and Affective Technologies}

Designing technology to support reflection is an important domain in HCI \cite{molsInformingDesignReflection2016a}. Fleck and Fitzpatrick discuss how technology can fall upon a spectrum of reflection, from broader revisiting and exploration to deeper transformative change and criticality \cite{fleckReflectingReflectionFraming2010}. They outline the importance of understanding the purpose of reflection in design, which behaviours are encouraged, and whether the context of time and space is suitable for reflective practice \cite{fleckReflectingReflectionFraming2010}. Bentvelzen et al. highlight four design resources that support reflection --- time, conversation, comparison and discovery \cite{bentvelzenRevisitingReflectionHCI2022a}. 

Starting with time, Hallnäs and Redström's \emph{slow technology} introduces a design paradigm focused on promoting reflection and mental rest through reducing the speed of interaction \cite{hallnasSlowTechnologyDesigning2001a}; a concept explored heavily in Odom's work \cite{odomPhotoboxDesignSlow2012a, odomAttendingSlownessTemporality2018a, odomInvestigatingSlownessFrame2019}. Odom highlighted how slowness in design can manifest in different ways --- it can be implicit or explicit, as a part of usage or as a part of the system \cite{odomExtendingTheorySlow2022a}. For example, design friction --- intentional hindrances --- can slow an interaction down to promote mindful usage \cite{mejtoftDesignFriction2019, coxDesignFrictionsMindful2016}. 

Extending slowness in design to communication technologies, King and Forlizzi \cite{kingSlowMessagingIntimate2007} emphasize how emotional responses in messaging are often tied to timing, effort, and context. Furthermore, Seo et al.'s BeeperRedux  \cite{seoBack1990sBeeperRedux2025} reintroduces a slower communication pattern from the past, highlighting how incorporating friction or complexity in this context can extend personal autonomy and emotional management instead of focusing purely on productivity and gratification. Emotional management ties into research regarding emotional regulation design, which explores how technology can mediate emotional expression \cite{slovakDesigningEmotionRegulation2023, liuRegulationSupportCentering2025} (such as using music to steer towards positive emotions \cite{smithDigitalEmotionRegulation2022}).

In a writing context, the appropriate temporality and mindset are important for supporting reflectivity \cite{angeniusDesignPrinciplesInteractive2023a}. Space, platform, time, and materials all play an important role in shaping a writing practice \cite{linCraftingPersonalJournaling2025}. For instance, journaling is affected by the affordances of the medium, such as its permanency and editability, which can affect people's reflective behaviours \cite{linCraftingPersonalJournaling2025}. These dimensions together form a \emph{ritual} centred around the act of writing \cite{angeniusDesignPrinciplesInteractive2023a}; resulting in outcomes such as remembrance and letting go (i.e. burning the memory). Structured rituals in design can offer symbolic representations of transformative actions \cite{sasDesignRitualsLetting2016}, shifting the emotional landscape. Thus, rituals are another important resource in reflective design --- they can help people make meaning through difficult phases of life \cite{wolfDidDigitalTidying2022}.

We position unsent messages as inherently reflective --- authors revisit, explore, and reframe their initial feelings. They are positioned uniquely in the temporal dimension of Walther's \emph{hyperpersonal} model, and represent an unstructured ritual with various purposes --- to remember, to let go, to hold. Like Lin et al. with paper journaling \cite{linCraftingPersonalJournaling2025}, we explore how changes in their platform design affect the ecosystem of unsent messages and their reflective and affective behaviour.

\begin{table*}[]
\centering
\Description{Table displaying demographic information (age, gender) as well as the relationship(s) with the target of the unsent message(s).}
\rowcolors{2}{white}{mypink2}
\begin{tabular}{C{1cm} C{2cm} C{2cm} C{8cm}}

\hline
\rowcolor{mypink2}
\textbf{ID} & \textbf{Age} & \textbf{Gender} & \textbf{Relationship(s) of Intended Recipient} \\
\hline

P1*  & 22 & Woman  & Friend        \\
P2*  & 26 & Woman  & Romantic Partner        \\
P3*  & 39 & Woman  & Romantic Partner, Family Member        \\
P4*  & 20 & Woman  & Romantic Partner        \\
P5*  & 27 & Non-binary  & Romantic Partner        \\
P6  & 19 & Woman  & Teacher        \\
P7  & 20 & Man  & Friend, Romantic Partner, Teacher, etc.  \\
P8  & 20 & Woman  &  Romantic Partner       \\
P9  & 27 & Woman  & Family Member        \\
P10  & 22 & Woman  & Teacher        \\
P11*  & 21 & Genderqueer  & Family Member        \\
P12  & 24 & Woman  & Romantic Partner        \\
P13*  & 21 & Woman  & Friend        \\
P14  & 59 & Woman  & Family Member        \\
P15  & 23 & Woman  & Romantic Partner       \\
P16  & 22 & Woman  & Romantic Partner         \\
P17*  & 28 & Woman  & Friend, Family Member        \\
P18  & 28 & Woman  & Romantic Partner          \\
P19  & 34 & Man  & Friend        \\
P20  & 21 & Woman  & Friend        \\
\hline
\end{tabular}
\caption{Summary of interview participants. * represents participants who took part in the follow-up probing study.}
\label{table:demographics}
\end{table*}

\section{Formative Study --- Exploring the Experience of Unsent Messages}

The goal of our formative study was to address \textbf{RQ1}. Although unsent messages resemble commonly studied reflective, writing-based practices such as journaling or diarying, there is a sparsity of academic research around specifically this practice, which carries a distinctly social element. Our formative study holistically explored the practice through listening to accounts of such experiences.

\subsection{Participant Recruitment}

To recruit participants, we posted on our institute's paid studies board. We provided a pre-screening questionnaire, asking if the respondents had experience with unsent messages and, if so, to provide a brief description of their experiences. We used collected responses to purposively sample participants representing a broad range and depth of experiences. We recruited 20 participants (average age: 26.2, ranging from 18 to 59; gender distribution of 2 men, 16 women, 1 non-binary, and 1 genderqueer; see Table \ref{table:demographics}). This sample exceeded local standards in HCI \cite{caine2016} and provided satisfactory information power \cite{malterudSampleSizeQualitative2016a}. We do note that our gender distribution is skewed towards women, which was also representative through the pre-screening survey. Although this could suggest a gendered effect of writing unsent messages or a tie towards broader discussion on the gendered effect of emotional self-disclosure \cite{snellMensWomensEmotional1989}, our data does not support further speculation on this point. Before the study, participants were asked to read, review, and sign a consent form regarding data collection and usage --- this study was approved by our institute's ethical review board. 

\subsection{Study Protocol}

The formative study involved a semi-structured interview conducted online over Zoom. Given that many participants disclosed that unsent messages were typically tied to sensitive, emotional moments, we reminded them that they could refuse or skip questions at any point --- the primary researcher took extra attention in being tactful throughout the study. We started by asking participants to expand on the context they had provided in the pre-screening questionnaire, before expanding with questions regarding how they felt, why they chose not to send it, what platform they used and why, and so forth (interview questions can be found in the supplemental material). The full user studies took between approximately 30--50 minutes, depending on the respondent and their experiences, and participants were compensated \$16 CAD / hour. 

\subsection{Data Analysis}

The data was analyzed through a reflexive thematic analysis approach \cite{braun2021TA}. To start, the primary researcher reflected on their positionality in the research --- being someone who has written such messages in the past in emotional moments or arguments --- thinking through their own reasoning or rationales for doing so. While conducting studies, the primary researcher constantly reflected on the data and noted their thoughts about certain participant responses, aiming to raise self-awareness regarding their own biases and experiences. After thinking through the studies and familiarizing themselves with the data, the researcher started with initial unstructured coding. Coding was not a linear process --- codes constantly evolved, shifted, and combined during the process. At the end of the process, the codes were evaluated against the data again, and we felt we had satisfied a critical level of analysis. During and following the coding process, the researcher considered relationships between our codes, grouping and aligning them towards broader labels that helped us develop our subsequent themes, presented in the findings. 
 
\subsection{Findings}

\subsubsection{Unsent Messages as Containers for Unspoken Feelings}

When participants considered the driving motivations behind their unsent messages, they shared anecdotes relating to how such messages held feelings, among others, 

of anger --- as in P7's case, towards many interpersonal relationships in their life:

\begin{quote}
    \emph{``Whether it's a friend, a girlfriend, whether it's a social worker or professor... sometimes I would just get really angry with them, like any other person, and I would consider sending [a] message.''}  - P7
\end{quote}

of guilt --- as in P11's case, for not reaching out to their grandparents in an ongoing tenuous family circumstance:

\begin{quote}
    \emph{``It's my guilt for not talking to them... I'm also worried that they'll pass before I'm able to talk to them and explain why I don't keep in contact.''} - P11
\end{quote}

of obligation or necessity --- as in P13's case, of setting boundaries with their friend:

\begin{quote}
    \emph{``With my best friend... I was getting into a relationship and... I started to set more boundaries about just being a friendship.''} - P13
\end{quote}

of loneliness --- as in P1's case, of wanting to message someone familiar in an isolated environment:

\begin{quote}
    \emph{``[I felt] lonely. I was having trouble making friends in university... I was like maybe I should ask how he's doing and try and get him to talk to me again''} - P1
\end{quote}

Initially, when participants started writing such messages, the vast majority mentioned that they \emph{had} intended to send them off. They hoped for a resolution, because they missed what used to be, or they wanted to continue the relationship. For example, even though P15's ex no longer messaged them, they \emph{``still want to be friends... so I wrote these lists of things just in case in the future, if he changes his mind''}; and P19 mentioned that writing a message asking someone out for dinner was accompanied by a sense of \emph{``excitement... like you're eager to know the response''}. For all cases, unsent messages held strong feelings that participants felt they initially wanted to convey. 

So, even after writing these messages, why did participants not send them off? This reluctance was primarily drawn from worry about the possible repercussions and outcomes. Participants were reluctant to demonstrate vulnerability and place themselves into positions that they might not be familiar with, citing prior experiences that shaped their personalities. For instance, P4 mentioned that growing up \emph{``I wasn't allowed to be angry or express any sort of distress... it was very much a taboo concept''}, and in writing an angry note towards their ex during a breakup, realized \emph{``I don't really like expressing any sort of vulnerability on my end. So it was a little jarring for me, to be so emotionally vulnerable and honest''} (P4). Withholding this vulnerability could help prevent hurt, as P12 and P1 internally debated whether to send their messages, ultimately deciding not to: 

\begin{quote}
    \emph{``What would make me feel better? Would it be an apology? What if he didn't [apologize]? What if he was like I don't care, why do you care so much?... I thought it was like protecting myself, to not send.''} - P12
\end{quote}

\begin{quote}
    \emph{I'd say... not wanting to be let down by someone, or be disappointed by the answer''} - P1
\end{quote}

Beyond self-perception, participants were scared of the outcome --- that speaking their true feelings might be a burden, \emph{``I don't want them to just be thinking... I remind them about [his death], because that's all the experience they have had with me.''} (P14) or that they might ruin a relationship, \emph{``I valued the friendship so much, I didn't wanna change it or make things weird.''} (P13). This leads us to the core of the practice --- \textbf{people write unsent messages to contain the feelings towards others that they feel they cannot say to them}, because of potential consequences.

\subsubsection{Unsent Messages Give Shape to Imagined Scenarios}

The vast majority of the feelings that drive unsent messages were negative --- emotions that participants already felt could be burdensome or received negatively. During and after writing, participants imagined how the other party might respond to their written messages containing these emotions and considered their further-reaching consequences. P18 mentioned that \emph{``I'm not saying the nicest things and I know that will harm the conversation and potentially hurt his feelings a lot''}; similarly, P2 mentioned that \emph{``I imagined that he would get pissed off... and because of that I was heavily editing it so he wouldn't''}. Yet, without sending the message, how exactly do participants mentally construct the consequences that they are worried about? 

Some participants had previously sent messages that expressed such feelings, and they felt that it ended up making things worse. For example, in circumstances where they were mad at their partner, P18 mentioned that: 

\begin{quote}
    \emph{``There was one or two times where I actually typed out a message and said some not nice stuff, sent it without thinking, like very impulsive[ly] and it caused a bigger argument.''} - P18
\end{quote}

While almost all participants connected consequences to consideration of the other party's feelings, P7 expressed a unique opinion that social consequences regarding their messages were a mark against the sensitivities of general society after they recounted situations in which authorities had gotten involved over what they had sent:

\begin{quote}
\emph{``Even if I don't think it's gonna be threatening, [or] it's gonna be mean-spirited, [or] it's gonna be defamatory, [or] it's gonna be violent... I feel like I have to walk on eggshells in today's world.''} - P7
\end{quote}

Contrasting such cases in which participants had experienced concrete consequences, there were also circumstances in which participants were scared of consequences constructed purely mentally. In many of these cases, participants recounted ``overthinking'' or spiralling into anxiety, over-fixating on the worst case. They became doubtful of others, facing uncertain consequences and unanswered questions:

\begin{quote}
    \emph{``I definitely, like I wrote it, I had already villainized him in my head. I had this perception of him, and I was like, `Oh, you don't care.' So it won't matter what I put down, you're not gonna care.''} - P4
\end{quote}

\begin{quote}
    \emph{``What if they don't want me on the phone? Or to email them, or chat with them on Messenger. Like, what if they're like, oh God, it's her? ... Not that they would, but then, that self-doubt... I don't know.''} - P14
\end{quote}

These specific anecdotes are particularly telling --- for P4, they mentioned that after meeting the recipient later, \emph{``he definitely would have cared. knowing what I know now, if he had read the letter, he would've cried''}. P14 repeated several times how strongly they believed that the recipient would want to reconnect, yet still felt doubt --- highlighting how overthinking can overwrite even strong faith in others. 

Regardless of how these consequences were formed or what they concerned, the basis was the same --- participants weighed their desire to share against their desire to avoid negative consequences, and always chose the latter. The imagined dialogue and outcomes highlight how unsent messages are deeply relational (contrasting, e.g., a journal entry), as they tether the feelings from the writer towards the recipient. Without consequences, P5 explained that they may be more willing to share their feelings: 

\begin{quote}
    \emph{``I hope maybe someone could find this, you know, if I die... But I don't want to be the one... I want people to know, people should know this ... I don't want to deal with the external chaos and pressure''} - P5
\end{quote}

\subsubsection{Unsent Messages Take On Ritualistic Qualities}
\label{sec:findings1ritual}

Yet, the prior two themes highlight a notable problem --- if unsent messages represent the strong feelings that one has towards a relationship (like anger, like hurt), and if those feelings are not necessarily expressed in the moment, then how might the relationship progress? In some cases, participants used unsent messages as a ritual to mark the conclusion of a relationship --- writing down, confirming their feelings, and then almost figuratively ``letting go''. P4 and P8 both described this in the case of a romantic breakup: 

\begin{quote}
    \emph{``The way that I've always coped with stuff like this is to write letters... So my plan was to just write this letter, send it to him, be done with it, and wash my hands of it and everything.''} - P4
\end{quote}

\begin{quote}
    \emph{``I wanted to just say what I wanted to say, and just get out of that feeling, and free myself.''} - P8
\end{quote}

To further emphasize the ritualistic nature of letting go, P8 drew an analogy towards how, when they were younger, \emph{``I used to write down on paper and just burn them, or flush them down, so it just felt better''}. 

Yet, not all unsent messages were intended to represent a parting moment. For others, unsent messages became a cyclical ritual of writing, feeling, and understanding the consequences amid a tangled relationship. P3 understood that expressing their thoughts during anger could be hurtful, and deliberately wrote unsent messages to cool down their emotions, understand their feelings, and possibly later communicate their words in a meaningful, constructive way: 

\begin{quote}
    \emph{``I'll take a moment, maybe 5 minutes away, and I'll go on my phone and pretend to write him a message about everything he did that pisses me off. And then I don't press send... When I actually have a moment... I'll go through the message.''} - P3
\end{quote}

The temporal transformation of unsent messages into a habitual process tempered the instinctive desires to send, and shifted them into more reflective exercises. The same ritual described by P18 was even encouraged by their partner to help them think through things and calm down. Yet, whereas these rituals tend to have positive reflective outcomes due to the participant's assuredness in understanding their wants and the potential consequences, the cyclical ritual of unsent messages could also cause participants to get ``stuck'' in their thoughts, where they are tugged between the various tensions of what they feel, what they want, and what they think they should do. For P11, this ongoing cycle made them still question what they should do regarding a message they wanted to send to a relative, as they had for years:

\begin{quote}
    \emph{``This cycle... I don't know what I would do if they responded positively to what I was saying... I feel like it's a lot easier to imagine what could go wrong and then that's my justification for not talking to them.''} - P11
\end{quote}

For P13, a similar negative cycle led to an eventual breaking point and the loss of a friendship:  

\begin{quote}
    \emph{``It was just kind of like a cycle. We'd hang out, a similar thing would happen, I'd want to send a text, and I was just thinking... what if he reacts badly, and then... yeah.''} - P13
\end{quote}

In all forms, as a deliberate and symbolic act or as a repetitive habit, we saw that unsent messages took on a ritualistic shape, giving a structured set of actions, a chosen space, and a moment in time to express and confront the feelings in the way that participants felt they needed to. 

\subsubsection{Unsent Messages Help with Processing and Reflection}
\label{sec:findings1processing}

For essentially all participants, even for the ones who seemed ``stuck'' in the moment, the writing process of unsent messages was a way of processing and reflecting on themselves and their relationships. P1 stated that it helps that \emph{``you've shared it with someone''}, even if that `someone' is yourself. Although reflective writing has been well-studied as a way to process, we discuss how specifically writing to share seemed to uniquely help. Firstly, the intention of the message and the internal debate between whether or not to send, caused people to consider their self-perception - how they wanted to present themselves to others (which was often not conducive to actually sending off what they had written). For P12 and P17, the messages would violate their cultivated self-image, which made them hesitate:

\begin{quote}
    \emph{``I like being nice and friendly and understanding... so I thought this message made me seem... kind of vulnerable because I was caring too much and I was hurt.''} - P12
\end{quote}

\begin{quote}
    \emph{``I felt it would ruin the self-image of the composed clinician or the level-headed therapist... I didn't want to come across in terms of self-image as being like the angry Asian woman.''} - P17
\end{quote}

Writing specifically towards someone also allowed for a second perspective on the relationship one had with another, which manifested both positive and negative perceptions. Positively, it allowed P3 to reconsider who their husband was beyond his immediate, anger-inducing actions; negatively, it allowed P13 to reconsider their relationship with their longtime best friend and whether it was worth maintaining anymore: 

\begin{quote}
    \emph{``When I start writing... it reminds [me], `Hey let's go back to square one, let's get grounded, let's think about who he was, or who he is, and why you're in this relationship... that he has empathy and compassion.'''} - P3
\end{quote} 

\begin{quote}
    \emph{``Having to think about how this is really affecting me. And I think typing it out or getting it into words helped me gain clarity about, `What is this friendship actually giving me?'''} - P13
\end{quote} 

Re-thinking the relationship also provided participants with insight about themselves as well. P5 and P15 both discussed how they learned about what they are actually looking for in a romantic relationship, and P8 discussed how they realized that past family trauma had coloured their perspective in all subsequent relationships. By spotlighting their true feelings through an unsent message, P20 mentions --- \emph{``It clears my mind. I can write everything without being scared of someone reading it, or someone judging me for what I'm writing... I can directly express how I'm feeling''}. All in all, the writing process of unsent messages helped both contain people's overflowing emotions, but also acted as a mirror for people to make sense of the situation that they were in. 

\subsubsection{Platforms Act as Containing Spaces for Unsent Messages}
\label{ref:platforms}

The choice of where the unsent message was written was deeply tied to the participant's intentions. For participants who initially had a strong resolve to send off the message, the message was written directly in the message box, keeping them close to the imagined interaction. However, the majority of participants discussed writing their unsent messages on a separate platform (n=12), telling us about how they were a bit more skeptical about whether to send it, wanted privacy away from the conversation, or somehow knew that they might need more time to process the message. This was essentially always a note-taking app (e.g. the default ones on a smartphone, or related apps such as Google Docs). Although these notes apps can differ slightly across phones or specific applications, various common features explained why participants used them:

\begin{itemize}
    \item \textbf{Privacy and Security}: The most common reason that a participant might write on an external app, like the Notes app, was for privacy and security. For P2, this represented \emph{``a very private area, I think it's just meant for you... Because it's private, it's a space for only you to connect your thoughts.''}. This provided a sense of safety to alleviate their anxiety --- \emph{``It's safer... No one's gonna see it.''} (P12). 
    \item \textbf{Distance from the Recipient}: Relatedly, the external app gave participants an increased sense of distance from the recipient. Part of this was because of the cues on how the recipient might be online or available, and they might read the message or it might be left as delivered or seen --- as P2 mentioned, \emph{``knowing that people are available, to be able to respond to something... fosters that anxiety in you''}. This increased sense of distance served a practical purpose as well --- to alleviate worry that they might accidentally click `send'.  
    \item \textbf{Record-Keeping}: The notes app keeps a longer record of the unsent messages, unless they are explicitly deleted. Although many participants felt that their written feelings were ephemeral, there were a few participants who wanted to keep and remember their feelings. P4 stated that \emph{``I can revisit certain ones and re-read it... and it's just good to reminisce sometimes''} and P7 mentioned that \emph{``I want to remember the feelings of anger without being angry... forgiving, but not forgetting''}. 
    \item \textbf{Structure and Organization}: Some participants explained that the notes app was more conducive towards organizing their thoughts in a structured way (tied towards processing and sorting their feelings). For P4, \emph{``It gives you more space, I can do headings and bullet points''}, and for P15, \emph{``I need to sort my feelings about it and make it a clear message first''}.
\end{itemize}

However, notes apps were not the only platform used --- 2 participants also mentioned writing such messages on pen and paper sometimes, being more visceral and symbolic (which P4 described as \emph{``feelings in the rawest form; because there's no revision, I know there's no changing it''}; and P5 described as being \emph{``more romantic''}). 

In general, participants' choice of platform for unsent messages might not have been given much conscious thought, yet the specific features and affordances of the platform were imperative in shaping how secure and contained participants felt. The notes apps helped participants articulate their feelings and process their situation, while maintaining a distance from the relationship; all contributing to the ritual of unsent messages as a whole. 

\section{Speculative Design Probes}

\begin{figure*}[h]

  \centering
  \includegraphics[width=0.95\linewidth]{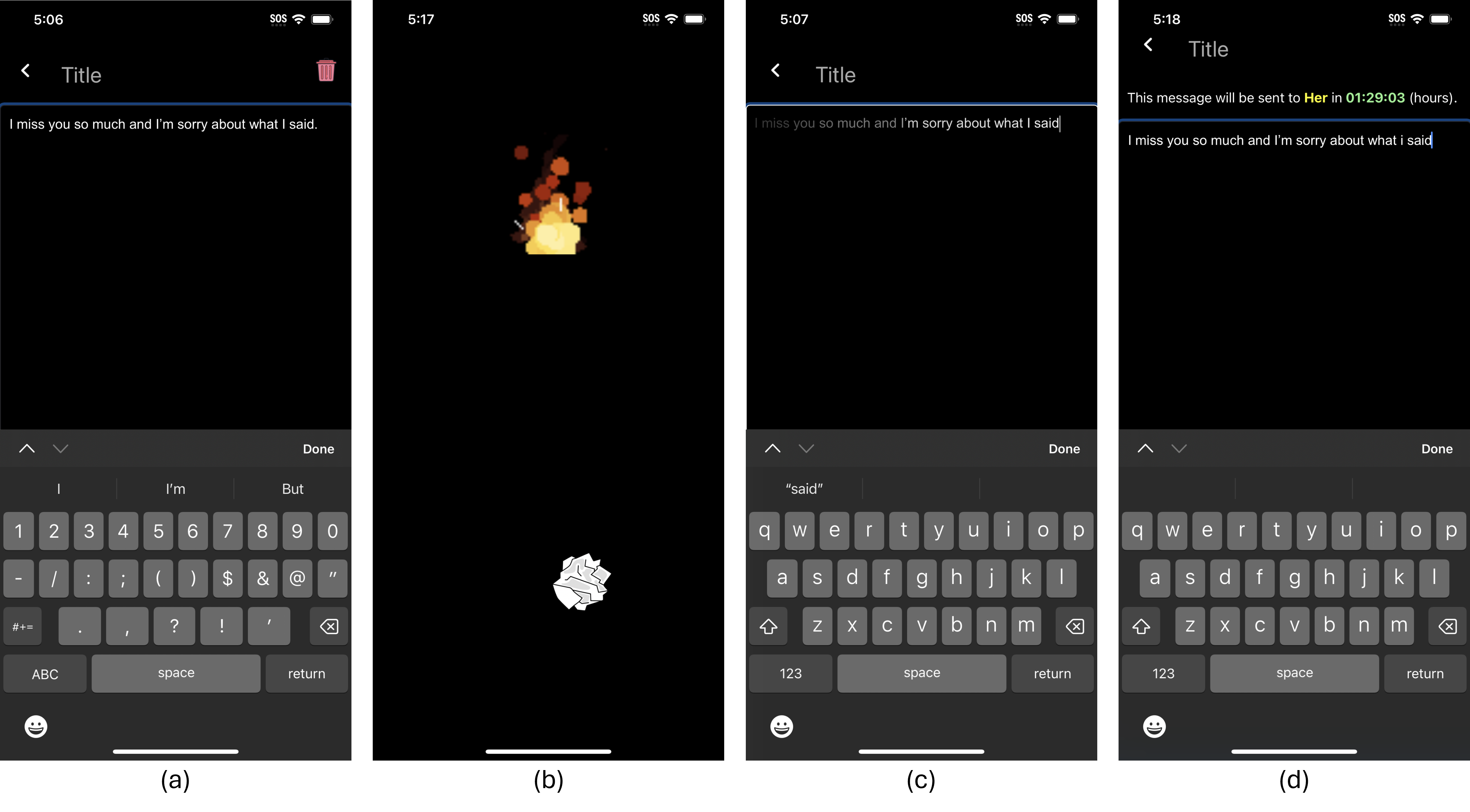}
  \caption{Some examples of our probes. Subimage (a) is our default --- an implementation of a basic notes app. Subimage (b) is Probe-Burn --- which invites the user to burn their written message after writing it. Subimage (c) is Probe-Fade --- in which messages fade over time and eventually get deleted as they are written. Subimage (d) is Probe-ForcedSend --- which imagines if unsent messages were instead sent automatically after a set time.}
  \Description{The image shows 4 subimages. Subimage (a) shows a default note-taking app. In the content of the note, the user has written an apology and an expression of emotion. The user can also add a title, return to the previous menu, or delete the note. Subimage (b) shows one of our probes, in which the user has crumpled their message into a digital paper, which they can toss into a digital fire. Subimage (c) shows something similar to (a), except now, the earlier characters in the message are fading out. Subimage (d) shows something similar to (a), except there is an added message - ``this message will be sent in XX hours''}
  \label{fig:hcd}
\end{figure*} 

Our formative study addressed \textbf{RQ1}, understanding the experience and motivations around writing unsent messages. Section \ref{ref:platforms} also provided initial insights into \textbf{RQ2}, by highlighting why people appropriated existing apps as improvisational spaces to write unsent messages. The design of the writing platform mediates the experience: the externality of the notes app made the space feel safer, and the automatic saving made the writing feel permanent.

In this subsequent study, we explore how deliberately altering the design of a platform affects how people feel when constructing unsent messages. With the initial basis of a note-taking app, we tweaked elements relating to affordances, rituality, sociality, and temporality to create 9 distinct prototype variants, which we speculated on with participants. 

Our methods drew inspiration from design probes, speculative design, and speculative scenarios, building on existing methods in Gulotta et al. \cite{gulottaDigitalArtifactsLegacy2013b}, Odom et al. \cite{odomTechnologyHeirloomsConsiderations2012}, and Lee et al. \cite{leeSpeculatingRisksAI2023}. We developed prototypes that aimed to be provocative --- exaggerating or subverting existing design decisions to invite imagination about how such interventions may change the existing experience of writing unsent messages. At the core, each prototype was based on our implementation of a familiar Notes app (Figure 1a); each variation explored a different dimension of the experience.

\begin{itemize}
    \item \textbf{Rituality} --- Unsent messages mimicked rituals, as a symbolic action taking place at emotional moments, using specific (software) tools for the purpose of handling emotions and driving change (Section \ref{sec:findings1ritual}). We were inspired by ritualistic design, in which patterns of actions are supported through symbolism, structure, and action \cite{sasDesignRitualsLetting2016, kluberDesigningRitualArtifacts2020b}. For instance, symbolic action has been studied as a means of completing emotional closure \cite{wagenerMoodShaperVirtualReality2024d}, we explored how adding symbolic actions might affect the experience --- \textbf{Probe-Burn} invited the user to ``burn'' the message by throwing it into a digital fire (Figure 1b). Appraisal of goals is also important for the actions taken during emotional regulation \cite{slovakDesigningEmotionRegulation2023}, \textbf{Probe-PreReflect} prompted appraisal through pre-writing reflection before writing a message in which users would need to answer why and what they wanted to get out of their writing. These answers would then be shown each time a user wanted to review their message. 
    \item \textbf{Temporality} --- Time was a consistent underlying thread throughout the themes - unsent messages involved participants taking time to write, reflect, cool down, and imagine the past, present, and future (such as in Section \ref{sec:findings1processing}). Inspired by temporality in design and its mediation of reflective outcomes \cite{odomExtendingTheorySlow2022a, odomPhotoboxDesignSlow2012a, bentvelzenRevisitingReflectionHCI2022a}, we experimented with probes that affected the temporal dimension of reflection. For instance, the emotional responses tied with the visibility of decay \cite{gulottaDigitalArtifactsLegacy2013b} inspired \textbf{Probe-Fade}, which entailed fading messages that disappear as you write (Figure 1c). Contrastingly, \textbf{Probe-Permanent} makes messages permanent, as you cannot edit them after saving. To explore the reflective value of future revisitation \cite{odomDesigningSlownessAnticipation2014a}, \textbf{Probe-Reminder} saved the message with a future notification to encourage re-reading of messages.  
    \item \textbf{Affordances} --- On the present platforms used for unsent messages, writers have a level of afforded freedom in control, as writers can deliberate over what they want to do with the message --- whether to send it, to reword it, to withhold, and so forth. We were inspired by counterfunctionality, which demonstrates that functional shifts can support new ways of thinking and interacting \cite{pierceCounterfunctionalThingsExploring2014d, seoBack1990sBeeperRedux2025}. We deliberately modulated an affordance inspired by privacy controls, with privacy representing an important interpersonal boundary during disclosure grounded by the decision criteria of the individual \cite{andalibiDisclosurePrivacyStigma2020}. To modulate the afforded decision-making freedom that authors typically maintain, we designed \textbf{Probe-ForcedSend}, which forced whatever was written to be sent after a set amount of time (Figure 1d), and \textbf{Probe-Unsendable}, which forced whatever was written to never be sent. 
    \item \textbf{Sociality} --- Section \ref{sec:findings1processing} shows that unsent messages inhabit a middle space in terms of sociality --- they represent the feelings that people want to share with someone else but could only share with themselves. We were inspired by the social aspect of reflection \cite{bentvelzenRevisitingReflectionHCI2022a}, and the use of AI in reflective technologies as a pseudo-social support tool \cite{kimDiaryMateUnderstandingUser2024b, kimDiaryMateUnderstandingUser2024c} to further explore this idea of ``private sharing''. For \textbf{Probe-SpeakBack}, the message was read by the system back to the writer (using OpenAI's \emph{tts-1} model), for \textbf{Probe-AIResponse}, an AI interlocutor read and offered suggestions back to the writer (using OpenAI's \emph{gpt-4o} model). 
\end{itemize}

These probes were intended to elicit imagined reactions and reflections rather than be fully functional tools. Even if functionality was not exactly enforced as intended (e.g. for Probe-Unsendable, we could not actually enforce written notes to never be sent), we found them sufficient to convey the intended interaction and invite discussion. For technical implementation, they were all developed using ReactJS and shipped as a progressive web app hosted on an online server. They are intended to be used with a mobile phone and support typing into the message box. Full screenshots and details of all the probes are in the supplemental material. 

\section{Probing Study --- Exploring the Design of Unsent Messages}

The goal of our probing study was to explore \textbf{RQ2}, speculating with participants on how the design of such a platform may affect the experiences of unsent messages. 

\subsection{Participant Recruitment}

Participants were recruited from a purposive sample of our formative study. We considered participants who had gone into experiential depth with rich, detailed accounts; we sent them an invitation for a follow-up. This represented a sample of 8 participants (average age: 25.5, ranging from 20 to 39; gender distribution of 6 women, 1 non-binary person, and 1 genderqueer person). Although a smaller sample compared to the prior study, we considered it adequate for an exploratory design probe, as we were more interested in preliminary breadth and depth rather than generalizability. We refer to participants by their ID from the formative study.

\subsection{Study Protocol}

Our probing study took place online over Zoom --- we opted for online studies for participant reach and convenience. We first briefly reviewed the experiences that participants had shared from the first study. We asked participants to use our default note-taking app to write how they would in that experience or a similar one, letting them get accustomed to our system. We then proceeded to explore the 9 probes using the same experience; the researcher shared their phone screen throughout to guide the walkthrough and demonstrate each probe's features. For each probe, we described it and showed an example to demonstrate its unique functions. We then invited participants to use the probe to write an unsent message. Afterwards, we asked open-ended, semi-structured questions regarding that probe. Some of these questions were common to all, e.g. \emph{``What was your first impression? How did it feel?''}, some were more specific to the probe itself, and some were general to the overarching dimension (i.e. rituality, temporality, affordances, and sociality). In general, our speculative design involved walking through and discussing both the present implementation and future possibilities. The list of interview questions can be found in the supplemental material. 

The order of probe dimensions was randomized for each participant, but the order \emph{within} each dimension was kept consistent for conversation coherence and continuity. As we did not have a direct view into the participant's screen, we relied on self-reports and system activity logs (such as keystrokes) to confirm engagement. User studies took approximately 75--90 minutes, depending on the respondent and their experiences, and participants were compensated at a rate of \$16 CAD / hour. 

\subsection{Data Analysis}

The data was analyzed through a thematic analysis approach, similar to the formative study. While conducting studies, the primary researcher constantly reflected on the data and the insights captured from each participant. After all data was collected, the researcher engaged in initial unstructured coding, identifying patterns and relationships that were then grouped and visually mapped. Using these groupings, the researcher developed relevant themes that were bounded yet holistically represented the data \cite{braun2021TA}. Although we inductively began with our identified dimensions and findings from our formative study as a basis for coding, we found that insights from our probes overlapped and were not easily untangled. Thus, our interrelated themes cut across our initial dimensions and represent what we learned about design as a whole, rather than addressing specific probes. These findings help us understand the impact of the \emph{design} of unsent messages, addressing \textbf{RQ2}.

\subsection{Findings}

\subsubsection{Authenticity, Self-Presentation, and Self-Performance Impacts Content}

The design of different probes affected the content of the writing, especially in how participants presented themselves in their unsent messages. In particular, participants discussed a spectrum ranging from writing their most honest, unfiltered, and emotional thoughts to their more polished, refined, and toned-down thoughts. 

Participants tended to express their more honest thoughts when the probes included ``safety'' mechanisms that prevented what they had written from being sent. These safety mechanisms were affordances that distanced the message from the relational other and made the message seem more ephemeral and less likely to be sent. For instance, messages from Probe-Unsendable felt \emph{``more honest... this is really how I feel''} (P10). This was true even when these safety mechanisms were purely mechanical and system-based. For example, a message that was symbolically deleted by the system \emph{felt} safer, even when participants could choose to delete their messages normally:

\begin{quote}
    \emph{``When you tell me that I can burn it, I just feel like it's a safe space... I can freely express because I don't know... I'm scared somehow someone would find it... But if I know it's gonna be burnt, I would be more brave in typing things out.''} - P5
\end{quote}

On the other hand, when the imagined audience had a stronger impression or the message felt more permanent, participants felt less safe in expressing themselves honestly. Instead, they were more filtered, cautious, and anxious about what they were writing. For example, in Probe-Permanent, P2 mentioned that \emph{``it would make me more anxious and cautious. I feel like knowing you kind of only have one chance to prove yourself''}. In these cases, participants elevated their self-presentation over their raw honesty. This even held when the imagined audience was \emph{themselves}, as evidenced by Probe-SpeakBack, which encouraged them to hear their own words. Several participants mentioned that hearing their words came off ``cringy'', ``cheesy'', or ``harsh'', which would cause them to rethink their message:

\begin{quote}
    \emph{``You're kind of able to imagine yourself in that other person's shoes, and then being told what you just wrote... It does sound very harsh.'' - P3
    }
\end{quote}

\begin{quote}
    \emph{``I was aware that I would be playing it back and so I was already writing things in a way that I was like... I will be hearing this out loud.''} - P2
\end{quote}

Participants discussed the importance of being honest with themselves in their unsent messages, as this helps them validate their feelings. However, even though these first thoughts are what they feel initially and are important to express, participants noted that raw emotion is not always what they want to express to others. Under this perspective, authenticity does not only represent static expression --- as P11 mentions, \emph{``I feel like they were both, I guess, authentic to myself. I wouldn't say one was like a fake version of me''}, even though they filtered themselves differently across different probes. On top of honesty, people also wanted to feel safe with those thoughts; thus, they also strongly felt the need to be presentable, filtering themselves both for others and themselves. Design can alter the feeling of safety and privacy, placing user feelings on a spectrum of authenticity between raw honesty to tailored ideality. 

\subsubsection{Holding on and Letting Go --- Different Behaviours Through Design}

While sending messages can represent a definite expression of your feelings, deciding to never send them is a representation of letting go; certain probes supported participants in letting go of their negative feelings through the perceived ephemerality of the message. This was most evident in Probe-Burn, in which participants were invited to burn their written entries. Participants described how the \emph{symbolic}, definitive act of burning gave them a feeling of finality and release, showing how symbolism in design can encourage a mindset shift even if the result is the same as regular deletion. 

\begin{quote}
    \emph{``It can be very cathartic... getting rid of something that you can't look at again, that it really is destroyed... I feel like when you burn something, it doesn't exist anymore.''} - P1
\end{quote}

\begin{quote}
    \emph{``It's the symbolism of it. It's like the act of, I almost want to say, the violent act of finality, as opposed to something that's quieter and more discreet. It's almost showier, and in a way, it sends its own message.''} - P4
\end{quote}

This symbolism was further evident in Probe-Fade and Probe-Unsendable --- participant mindset was affected by the behaviour of the platform. For Probe-Fade, P13 mentioned that \emph{``I lose it in my memory... it's like fading from my mind''}; for Probe-Unsendable, P4 mentioned \emph{``this ties back into the safety I was talking about... I think there is comfort''}. When participants felt like the platform spotlighted the ephemerality of their feelings --- that their written words could never be sent or saved, participants felt a sense of finality, and that they should let go as well. This further encouraged people to write more freely, tying back to the previous section. 

Even though letting go of negative feelings was typically a cathartic feeling for participants, holding onto their feelings could also be an important positive form of remembrance and self-honour. In opposition to the prior example of Probe-Fade as letting go, P2 stated that fading messages could be bad \emph{``in situations if you haven't fully processed your emotions, and there is a way for you to do that''} --- highlighting the value in retaining one's emotions to understand them. Participants wanted to re-read and reflect on their writings as a way to express themselves, remember the learnings and takeaways from their mistakes, and understand who they had been.   

\begin{quote}
    \emph{``I accept who I was... This is weird, but it's also a beautiful letter in my opinion that I record... I must be very, very sad or feeling really, really like I \emph{have} to write things down''} - P5, talking about wanting to record their writings when discussing Probe-Fade
\end{quote}

However, having the platforms hold onto their feelings for too long could be dangerous as well, as it could keep people stuck in their emotions, representing a cyclical loop similar to Section \ref{sec:findings1ritual}. For instance, P2 mentions that for Probe-PreReflect, which keeps around one's reflective questions: 

\begin{quote}
    \emph{``If you're the kind of person to hold a grudge forever, this would probably not be your thing, because I feel like it would remind you every time why you feel a specific way about a person. So it would be not conducive to letting go in some cases.''} - P2
\end{quote}

Overall, we find that participants valued holding and sitting with their lived experiences, even the negative ones, as a way of honouring their feelings and who they were. Yet, they also did not want to live in those moments forever, wanting to let go of them when appropriate. Design affordances, such as whether the unsent message was saved or not, altered how participants viewed the tension between holding on and letting go. 

\subsubsection{Supporting the Emotional Versus Reflective Texture of Unsent Messages}

We found that some probes encouraged more impulsive venting of people's immediate emotional feelings, and some probes encouraged more thoughtful contemplation of what to write in the message and why. From our formative study, we found that participants typically engaged in both --- they initially vented impulsively, and slowly started to engage in reflection about the context, their wants, and their relationships as they continued. Most relevantly, Probe-PreReflect flipped the usual temporal sequencing, encouraging people to think of the outcomes and consequences before they started writing. Participants described this as both impacting what they wrote and how they felt while writing. For instance, P4, who had previously deliberated on how to communicate with her ex for the last time: 

\begin{quote}
    \emph{``I think it channelled my emotions the best or helped me control them the best... I never really considered how I would like to feel post-writing the letter, and having that goal set affected how I wrote, because instead of writing blindly with emotion, it's now with intent.''} - P4
\end{quote}

By answering the ``why'' of writing, participants were able to set goals for themselves through their writing, providing them with direction in their words. However, participants also worried about whether they were not able to meet their goals, e.g. \emph{``it could set up expectations that can be disappointing''} (P3). Furthermore, not everyone \emph{desired} a deep level of reflection --- for instance, the goal of P5 in writing unsent messages was simply to express rather than to necessarily learn, tied to their consistent motivation throughout both studies in \emph{expressing} their honest feelings and memories:

\begin{quote}
    \emph{``I just want to write it when I see a paper or I really want to get it out. I don't have an expectation towards my actions... My emotions are really intuitive... I just feel [Probe-PreReflect] is more like a writing task at school than more emotional.''} - P5
\end{quote}

From Section \ref{sec:findings1processing}, reflection was supported through a second opinion on the participant's thoughts. Getting a second perspective on one's thoughts was most deeply supported by Probe-AIResponse, which offered an AI opinion on one's written message. Even though several participants found the AI opinions somewhat generic, most mentioned that it was good to receive validation and reassurance, e.g. \emph{``it was just saying I was valid, which is still nice''} (P11), \emph{``it's like you have a friend that isn't going to come and judge you unprompted''} (P2). However, even beyond that, AI could help with processing and understanding:

\begin{quote}
    \emph{``If potentially you are a person who struggles with naming emotions or feeling certain things... then you need to talk about it somewhere to somebody''} - P2
\end{quote}

\begin{quote}
    \emph{``It's asking the right questions about `how are you feeling about everything? Have you found ways to process your feelings?'''} - P17, drawing from their psychology background 
\end{quote}

Altogether, depending on the emotional and situational context, participants both desired initial catharsis and immediate release, yet also wanted to ponder on the meaning of the situation to take transformative steps. The former ties to impulsive venting that prompted writing; the latter ties to patient, slower reflection that came afterwards. Design can nudge in either direction, but we highlight that both are important in different contexts as ways to feel better. 

\subsubsection{Tension Between System Decision-Making and Personal Autonomy}

In contrast to typical note-taking apps, several of our probes add design elements that remove control from the user and pass it on to the platform. This was most evident in Probe-ForcedSend and Probe-Unsendable, which squash the decision-making of the user into an enforced binary choice in which the message is sent or never sent, respectively. Participants mentioned that by allaying the choice, the platform took on the responsibility for them as well. For instance, P2 mentions that \emph{``[Probe-ForcedSend and Probe-Unsendable] takes away the anxiousness of the unknown''}, and that:

\begin{quote}
    \emph{``If you're a wishy-washy person, like me, then I feel having the decision made for you would personally take away a lot of the anxiety...  if a person is more straightforward, is not swayed by decisions... or has a stronger moral compass... they would probably be OK making their own decisions.''} - P2
\end{quote}

Probe-ForcedSend in particular induced feelings of artificial pressure, e.g. \emph{``it is a little more stress inducing... I overthought the wording more''} (P13), \emph{``I feel like I would just be ruminating on the wording''} (P17). Even though this pressure felt uncomfortable for almost all the participants, some participants agreed that this could also be useful as a final push to prevent people from getting stuck in the cycle of indecision and feel better afterwards:

\begin{quote}
    \emph{``Although I'll be stressed in the [time limit], I think once it's done, that stress might be relieved a bit.''} - P13
\end{quote}

Yet, participants also mentioned the negatives of giving away this responsibility as well. Having the choice in their own hands could sometimes paralyze them, but working through their decision --- weighing what they valued, their feelings, and so forth, even when it feels awful --- was also imperative to personal growth. 

\begin{quote}
    \emph{``I feel like it's very much part of the human experience... I feel like it's very important in growing and maturity? And growing in understanding what kind of relation I want to have, how I want to treat others, how I want others to treat me.''} - P1
\end{quote}

\begin{quote}
    \emph{``Writing is a huge process of trying to understand. I think it's important. To understand your emotions and identify them and acknowledge them.''} - P3
\end{quote}

Participants leaned towards the freedom of choice, even when having choice can seem overwhelming. Participants admitted that they needed to sit in this discomfort to grow: P1 mentions that \emph{``to avoid feeling awful, I think you have to feel awful first''}. Still, participants also appreciated nudges by the platform when appropriate, to prevent themselves from being stuck in negative thoughts --- in this way, the platform helped contain such emotions.

\section{Discussion}

In this section, we revisit the significance of addressing our research questions. Firstly, we consider \textbf{RQ1} --- surfacing the implications of what we learned about the experience of unsent messages, their motivations, and the reflective impact. Then, we consider \textbf{RQ2} --- highlighting how intentional design impacts the entirety of the experience. 

\subsection{The Importance of the Liminal Space: Unsent Messages as a CMC-Specific Medium for Reflection}

Ultimately, we do not perceive unsent messages as a failure of communication, but rather a way of regulating conflict --- which Roloff and Ifert \cite{roloffConflictManagementAvoidance1999} state \emph{can} be successful in certain contexts. Unsent messages balance both people's authentic wants --- of sharing, of being honest, of seeking belonging and closeness \cite{bar-talSelfCensorshipSocioPoliticalPsychologicalPhenomenon2017, davisFriendship20Adolescents2012} --- against perceived risks --- of violating principles, of causing arguments, of destroying a certain self-presentation both for themselves and others \cite{madsenSelfcensorshipInternalSocial2016, andalibiDisclosurePrivacyStigma2020, bar-talSelfCensorshipSocioPoliticalPsychologicalPhenomenon2017}. The imagined other is important --- people do not know how the other person will respond, especially without cues \cite{walther2011theories}. In Walther's hyperpersonal model, this causes people to idealize their perceptions of the receiver and attribute certain stereotypical qualities \cite{waltherComputerMediatedCommunicationImpersonal1996d}. Yet, our findings suggest participants were aware of potentially idealizing into a positive relationship and almost overcorrected --- they had an idea of how the other person might react. They sometimes did not believe themselves, hedging against the worst case instead. In preparation for the worst case, people wrote unsent messages --- optimizing their presentation as they rehearsed \cite{waltherComputerMediatedCommunicationImpersonal1996d, waltherSelectiveSelfpresentationComputermediated2007} against an uncertain reading of the appropriate boundary and social context \cite{kennedyOversharingNorm2018}. 

From a social media standpoint, these actions were akin to defining the audience \cite{dasSelfCensorshipFacebook2013} and constructing the self-presentation in preparation for potentially vulnerable disclosure \cite{andalibiDisclosurePrivacyStigma2020}. Yet, we found that participants sometimes struggled with these points --- being unsure of how to disclose and present their honest feelings through communication, in fear of appearing vulnerable or being hurt. Extending the hyperpersonal model with this concept of vulnerability, we find that the theorized effects of CMC --- of idealization of the other, of struggle with self-presentation --- can also result in the distinct maladaptive emotional regulation strategy of \emph{catastrophization} \cite{zsidoRoleMaladaptiveCognitive2021}. Given this messy entanglement --- between people's inherent wants versus what they want to present, between not wanting to be hurt by the response versus wanting to be accepted --- a \emph{liminal} space in-between everything was born. 

Even though unsent messages were initially a way for people to handle their feelings towards another when direct disclosure felt risky or uncomfortable, they transformed into sites of reflection. Because CMC affords distance between interactors and non-instantaneous conversations, it represents a level of implicit slowness \cite{odomExtendingTheorySlow2022a} in conversation, where people can communicate synchronously but also choose the pace of conversation. Slowness gives rise to reflection \cite{hallnasSlowTechnologyDesigning2001a}, and participants were able to take the time during writing to pause, think, and reconsider whether they should send their message. While writing, participants cognitively processed and confronted their inhibited feelings \cite{baikieEmotionalPhysicalHealth2005a}, mimicking both the behaviour and outcomes of journaling \cite{ullrichJournalingStressfulEvents2002c} and expressive writing \cite{baikieEmotionalPhysicalHealth2005a, pennebakerWritingEmotionalExperiences1997a}, albeit in an interpersonal context. Over time, both during individual instances of writing or when unsent messages were transformed into reflective habits, participants made a fundamental shift in how their desires to send were regulated, choosing to withhold their messages despite their impulses to send them. This transformative shift in final actions recalls Fleck and Fitzpatrick's transformative level of reflection \cite{fleckReflectingReflectionFraming2010}, positioning unsent messages as inherently reflective. As people work through the \emph{ritual} of thinking, writing, reflecting, and withholding, they make meaning and act accordingly regarding the relational context and their personal feelings.

\emph{Unsent messages} represent a reflective practice that materialized naturally through people's journeys with social media and navigating interpersonal experiences. It was never necessarily designed for; rather, it emerged naturally from people's feelings and the platforms they have been given to express such feelings. By writing in an imagined letter, people \emph{choose} to sit in discomfort and navigate difficult choices to understand themselves. In naming this practice, we lend legitimacy and meaning in two main ways. Firstly, \textbf{we spotlight an internal, invisible struggle that often has no apparent manifestation} --- participants' unsent messages often lie private, and sometimes their disclosure to the researcher as part of this study was mentioned as the first time they had been shared. Understanding people's \emph{emotional needs} is important, and we highlight diverse strategies in which people deal with their emotional relational labour without disclosure (e.g. letting go, holding on, reflecting). Secondly, by understanding this experience fraught with discomfort, \textbf{we can explore how to expand existing improvisational spaces through intentional design} to support people's emotional needs \cite{slovakDesigningEmotionRegulation2023}, potentially offering \emph{safe} \cite{pearce2022your} discomfort and reflection.

\subsection{Expression to Reflection to Autonomy: The Structural Design of Unsent Messages}

\begin{figure*}[h]
  \centering
  \includegraphics[width=0.95\linewidth]{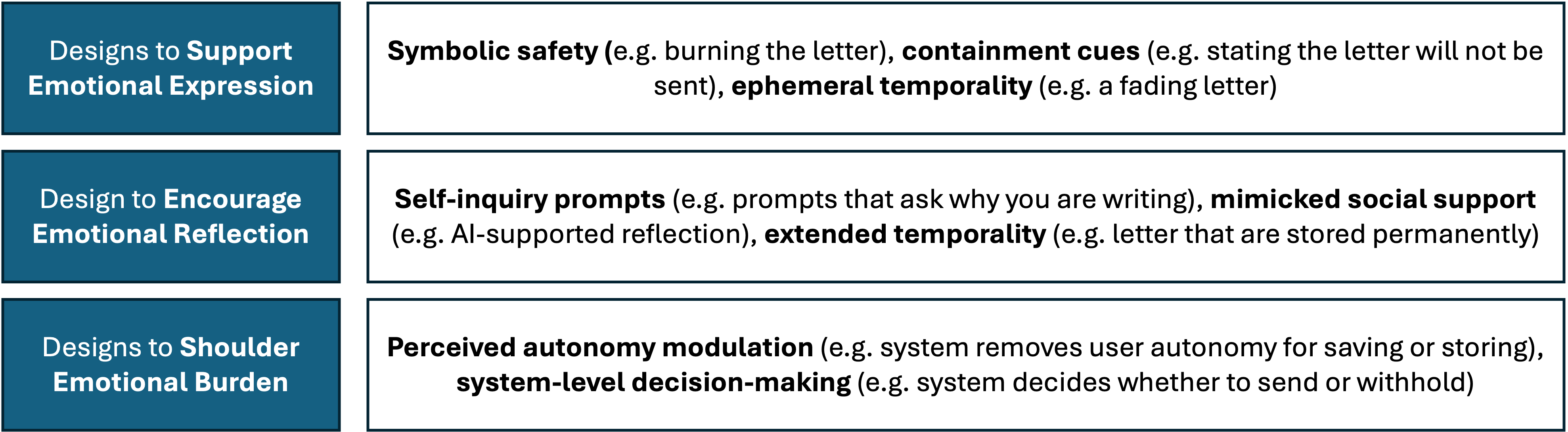}
  \caption{The emotional outcomes of unsent messages and the design features that nudge towards each outcome.}
  \Description{A figure showing three rows of emotional outcomes. For the first row - designs to support emotional expression, design features include symbolic safety, containment cues, and ephemeral temporality. For the second row, designs to encourage emotional reflection, design features include self-inquiry prompts, mimicked social support, and extended temporality. For the last row, designs to shoulder emotional burden, design features include perceived autonomy modulation, and system-level decision making. }
  \label{fig:outcome}
\end{figure*} 

Our probing study revealed that intentionally designing the platform for unsent messages affects how people react, form goals, and reflect on the outcomes of writing. Understanding design is important; with the same improvised space, outcomes were unintentional and varied. Some participants felt catharsis and reflection; others felt stuck in place. Design modulates the varied ways of emotional handling --- certain probes promoted letting go of feelings, others promoted holding on. Technology-mediated systems for writing and recording can promote positive outcomes \cite{isaacsEchoesHowTechnology2013a}, but exploring intentional design (e.g., as in \cite{seoBack1990sBeeperRedux2025, pierceCounterfunctionalThingsExploring2014d, odomPhotoboxDesignSlow2012a}) mediates the motivations and outcomes of such systems. We find that the platform design was able to shape a triad of emotional goals --- \emph{supporting expression}, \emph{encouraging reflection}, and \emph{shouldering burden} (Figure \ref{fig:outcome}) --- highlighting outcomes that designers can support.

Honest self-expression is important for authenticity \cite{kokkorisExpressRealYou2014}, tying into psychological need satisfaction and emotional well-being \cite{al-khoujaSelfexpressionCanBe2022}. Although human inauthenticity may not be the crux of the problem, its normalization can result in emotional states that feel undesirable and uncomfortable \cite{ericksonImportanceAuthenticitySelf1995}. Our findings show how design can shift the contexts in which participants \emph{feel} safe to express and confront themselves honestly in unsent messages --- in particular, designs in which their ephemeral emotions were symbolically faded, locked, or burned by the system. Symbolic digital safeguards allow users to sit with their most vulnerable, raw, and messy moments --- they need to \emph{feel} that their reflective practice is only for them. We underscore the importance of the symbolism and ephemerality in design as a way of generating digital safe spaces, which can promote honesty in expression and meaning-making \cite{heDesigningSocialComputing2009, eytamAssessingSubDimensionsSymbolism2023, rasch2024mindmansion}. 

The honest expression of raw feelings was cathartic, allowing for immediate mood improvement. However, while immediate self-expression spurs initial understanding, slowing down design to allow participants to reframe and review perspectives on the relationship and themselves led to deeper reflection \cite{fleckReflectingReflectionFraming2010}. Certain design features were able to mediate cognitive transformation and processing --- both the use of AI support in Probe-AIResponse and the pre-writing questions in Probe-PreReflect potentially support figuring out broader wants beyond just putting down their impulsive thoughts; instead of just writing continuously, these serve as design frictions to induce more mindful interaction \cite{mejtoftDesignFriction2019}. However, there are important ethical considerations as well, especially in delegating mental support to an external AI tool \cite{carrAIGoneMental2020, burleyEthicsAutomatingTherapy2024}. Adding support tools to support self-discovery and providing simulated social support can shift towards a more introspective mindset beyond an initial release of catharsis, agreeing with prior research \cite{bentvelzenRevisitingReflectionHCI2022a, liuRegulationSupportCentering2025}. We emphasize that such feelings are not opposed to honest expression: authenticity is not simply about the expression of raw feelings, but also about what people do afterwards to learn and transform. 

Several of our probes fundamentally affect the existing affordances of unsent messages, for instance, being forced to send it (Probe-ForcedSend) or never send it (Probe-Unsendable); not being able to permanently record one's feelings (Probe-Fade) or not being able to edit them after writing (Probe-Permanent). These designs thwart user autonomy, yet were sometimes beneficial in removing emotional burden --- unbounded freedom could sometimes leave people in emotional overload and stuck in place. Still, autonomy is important for human motivation; it is one of the core needs that Deci and Ryan propose in self-determination theory (SDT) \cite{ryanSelfDeterminationTheory2023, deci2012self}. Participants indicated that working through instances of uncertainty was imperative in learning about themselves. Discomfort (that comes with autonomy) is important for learning and growth \cite{kellerBecomingComfortableUncomfortable2022, zembylasDiscomfortingPedagogiesEmotional2012, pearce2022your}, yet there was a subtle line between sitting with the feelings and drifting into overwhelm. We find that the design of systems places autonomy along a spectrum, where system-autonomous designs can support and nudge users in certain (potentially helpful) directions. Yet --- there is a balance; given too much autonomy, people can spiral into a negative loop; given too little autonomy, people may delegate away the uncomfortable yet \emph{human} parts that come with reflection. 

Given people's diverse perspectives towards unsent messages, there is no one-size-fits-all solution. Participant narratives of unsent messages fall along several strategies of emotional management, from situation selection (i.e., avoiding conflict by choosing to write) to response modulation (i.e., writing to keep negative emotions in check) \cite{smithDigitalEmotionRegulation2022}. Furthermore, participants wrote for both hedonic and instrumental goals, aiming to both change their own feelings and seek broader understanding \cite{smithDigitalEmotionRegulation2022}. Still, understanding how design impacts emotional outcome has practical benefits. One relevant outcome is in understanding how the design of unsent messages can manage their associated emotional labour, which can negatively impact mental health. Unsent messages are a representation of a relational want, and for some, difficulties both in letting go and communicating directly led to an overwhelming cycle of rumination; the maladaptive foil of reflection \cite{watkinsReflectingRuminationConsequences2020, takanoSelfruminationSelfreflectionDepression2009, eikeySelfreflectionIntroducingConcept2021}. The design of a presently improvised space can therefore help, e.g., through social support, symbolic letting go, or taking on added autonomy to achieve resolution. By providing the relevant nudges --- \textbf{intentionally designed spaces can support and regulate people's emotional behaviours, helping them not slip into overwhelm}. 

\section{Limitations}

Our sample demographic skewed towards young women. This aligns strongly with previous research around disclosure and the potential gender effect \cite{andalibiDisclosurePrivacyStigma2020} and could highlight writing unsent messages as being more common to a specific demographic. Still, we highlight that future work could collect a diverse sample in age and gender, exploring how such dimensions might affect the motivations around unsent messages, e.g. exploring why certain demographics might be more likely to write such messages than others. Furthermore, we note that recruitment was largely local to a large urban centre in North America. Regional and cultural attitudes around communication styles may also play a role in our study (e.g., \cite{merkin2009cross}), further emphasizing the need for a broader sample. 

The thematic analysis was largely done by a single researcher. Although generalizability is not the goal of reflexive thematic analysis, which is subject to the lens of the researcher's perspective and lived experiences, other perspectives could have been beneficial in aligning and triangulating the presented themes. 

Finally, we reiterate that our work, especially our probing studies, was exploratory. Future work can take the learnings from our speculative design and translate them into functional and usable notes-based prototypes. These can be explored via longitudinal studies with a wider population to more broadly examine the roles of unsent messages. As a speculative direction, it would also be interesting to explore the relationship between unsent messages and mental health, as several related concepts (e.g., conflict management, catastrophizing, and overthinking) have theoretical implications on mental health \cite{zsidoRoleMaladaptiveCognitive2021, jamshaidOverthinkingHurtsRumination2020, deviExaminingRelationshipSleep2025, bearGenderEmotionalExperience2014}. Detailed exploration of participants would also help to understand their existing positions around self-disclosure and self-censorship, and highlight potential links to the goals of unsent messages. 

\section{Conclusion}

Our work looked at unsent messages --- messages that people write to others but end up not sending. Through formative interviews, we found that people wrote such messages in strong emotional moments, yet were worried about the possible repercussions of their words. The simple, incidental act of writing and then not sending organically became a container for people's feelings --- of the words they wanted to say and the scenarios that they imagined would happen. They took on a reflective nature driven by rethinking and second perspectives, and a ritualistic nature driven by time, space, and platform. In our probing study, we explored what would happen if we varied features of people's most common writing platform --- a notes-writing app. We found that design can affect people's authentic expression, their feelings of letting go or holding on, and their reflective outcomes. We discuss the implications of such designs, and speculate on how much design can and \emph{should} encourage reflective outcomes, support emotional venting, and help with holding difficult emotions --- letting people sit with and process their negative feelings by themselves.


\begin{acks}
This work was supported in part by the Natural Science and Engineering Research Council of Canada (NSERC) under Discovery Grant RGPIN-2019-05624. 
\end{acks}
\bibliographystyle{ACM-Reference-Format}
\bibliography{sample-base}


\end{document}